# Tuning the Electronic and Optical properties of Graphene and Boron-Nitride Quantum Dots through Molecular Charge-transfer Interactions

Arkamita Bandyopadhyay [‡, a], Sharma SRKC Yamijala [‡, b] and Swapan K Pati [a, c, *]

**Abstract:**

Spin-polarized first-principles calculations have been performed to tune the electronic and optical properties of graphene (G) and boron-nitride (BN) quantum dots (QDs) through molecular charge-transfer using Tetracyanoquinodimethane (TCNQ) and Tetrathiafulvalene (TTF) as dopants. From our calculations, we find that the nature of interaction between the dopants and QDs is similar to the interaction between the dopants and their two-dimensional counter parts of the QDs–namely, graphene and hexagonal boron-nitride sheets. Based on the values of formation energy and distance between QDs and dopants, we find that both the dopants are physisorbed on the QDs. Also, we find that GQDs interact strongly with the dopants compared to the BNQDS. Interestingly, though the dopants are physisorbed on QDs, their interaction lead to a decrement in the HOMO-LUMO gap of QDs by more than half of their original value. We have also observed a spin-polarized HOMO-LUMO gap in certain QD-dopant complexes. Mülliken population analysis, Density of states (DOS), projected DOS (pDOS) plots and optical conductivity calculations have been performed to support and understand the reasons behind the above mentioned findings.

## Introduction

Since its experimental discovery in 2004,[1] graphene has become as a potential candidate for application in nano devices because of its unique properties.[2-8] Similar to graphene, its inorganic analogues, namely, Boron-Nitride (BN), transition metal-dichalcogenites ($MoS_2$, $WS_2$, $VS_2$ etc.) and Silicon/Germanium 2-D sheets, have also emerged as interesting candidates in the field of nano-research.[9-12] Whereas pure graphene is a zero band-gap semiconductor or semi-metal, its inorganic analogues are either semiconductors (e.g. $MoS_2$[10, 12]) or insulators (e.g. BN sheet[11]). Thus, unlike graphene, these inorganic materials can be used in opto-electronic devices, without any modification.

Apart from its inorganic analogues, graphene's low-dimensional sisters, namely, 1-D nanotubes, 1-D nanoribbon and 0-D quantum dots (QDs) can also be used directly in opto-electronic devices because of their intrinsic band-gap.[13-17] As expected, the low-dimensional sisters of inorganic analogues also possess band-gap[18] and it has been shown that they can act as half-metals,[19, 20] cathode materials for rechargeable magnesium-ion batteries[21] etc. under the respective environments. Band-gap and other electronic properties of these low-dimensional materials depend on several factors. For example, band-gap of a graphene nanoribbon has been shown to depend upon its width, passivation and edge geometry.[17, 22-24] Similarly, graphene quantum dots (GQDs) have been shown to possess unique electronic properties depending upon their size,[15, 18] edge passivation,[25] and even, on their shape.[16, 25] Also, their tuneable energy gaps made GQDs as promising candidates for utilizing them in solar cells[26] and in LEDs.[27] Thus, generation of the band-gap and the ability to tune this band-gap are two important aspects for opto-electronic device applications.

Among the several ways of tuning the band-gap, molecular charge-transfer is proved as a simple and successful method, both experimentally[28] and theoretically[29-31] for fine-tuning the band-gap of graphene,[28, 29] its low-dimensional sisters[30] and also for its inorganic analogues.[31] In this method, based on the electronic donating/accepting nature of the dopant molecules, the substrate (i.e. graphene or its analogues) can be either n- or p-doped through the molecular charge-transfer interactions and these interactions will generate band-gap in the complexes (substrate + dopant).[28-33] Depending on the requirement, different organic molecules are used for doping. But, two molecules which were adapted for doping graphitic materials are Tetracyanoquinodimethane (TCNQ) and Tetrathiafulvalene (TTF),[28-33] where, the first molecule is an electron acceptor and the second molecule is an electron donor.[30] The electron accepting nature of TCNQ is because of its four cyano groups and the electron donating nature of TTF is because of the lone-pair of electrons present on its sulphur atom.

Though molecular charge-transfer method had succeeded in tuning the band-gap of carbon-nanotubes and graphene, yet, there is no report (neither theoretical nor experimental) regarding the usage of this method in tuning the band-gap of either GQDs or their inorganic analogues. In this study, using the first-principles calculations, we have tuned the band-gap of quantum dots [both Graphene QDs (GQDs) and Boron-Nitride QDs (BNQDs)] by doping them with TCNQ and TTF. From the studies, we find that the *dopant-QD interaction* is of charge-transfer type. To understand the dependency of the *dopant-QD interaction* on the size of quantum dots, we have considered three different sizes of QDs (both for GQDs and

BNQDs). We have also calculated the amount of charge-transfer between the dopants and the quantum dots. From all the results we understood that, both the dopants are physisorbed on the QDs and they can be used to tune the band-gap of QDs, and hence, are suitable for the applications in opto-electronic devices.

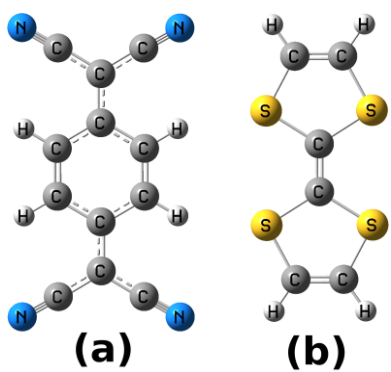

Figure 1: (a) Tetracyanoquinodimethane (TCNQ) and (b) Tetrathiafulvalene (TTF) molecules.

## Computational Details

All the electronic and optical properties of the systems have been calculated using the density functional theory (DFT) method as implemented in the SIESTA package.[34] Generalized gradient approximation (GGA) in the Perdew–Burke–Ernzerhof (PBE) form[35] has been used to account for the exchange-correlation function. Double-$\zeta$-polarized numerical atomic-orbital basis sets have been used for the C, N, S and H atoms. Norm-conserving pseudo-potentials[36] in the fully nonlocal Kleinman–Bylander form have been used for all the atoms.[37] A mesh cut-off of 400 Ry is used for the grid integration to represent the charge density. All the structures are considered to be optimized if the magnitude of the forces acting on all atoms is less than 0.04 eV/Å. As the systems are zero-dimensional, all the calculations are performed only at the gamma ($\Gamma$) point of the Brillouin zone. A vacuum of 20 Å has been maintained in all the three directions to avoid any unwanted interactions between the systems and their periodic images.

## Results and Discussions

**Structure and Stability**

In this work, we have considered both graphene and boron-nitride quantum dots of rectangular geometry and studied the effect of organic molecular doping to them. These rectangular quantum dots have a zigzag edge along their length direction and an armchair edge along their width direction. Following the convention of the graphene nanoribbons,[22] we represent these QDs as (21, 8)-QDs. Here, 21 and 8 are the number of atoms along the zigzag-edge (length, ~2.2 nm) and the armchair-edge (width, ~1.8 nm), respectively. Tetracyanoquinodimethane (TCNQ) and Tetrathiafulvalene (TTF) (figure 1) have been considered as the dopants.

Table 1: Energies of the complexes in different spin states (with respect to their ground-state) are given.

| Molecule | AFM (meV) | FM (meV) | NM (meV) |
|---|---|---|---|
| GQD-TCNQ | 0 | 20 | 750 |
| GQD-TTF | 0 | 90 | 770 |
| BNQD-TCNQ | 10 | 10 | 0 |
| BNQD-TTF | 10 | 260 | 0 |

From the previous studies on the interaction of TCNQ and TTF with graphene[29] and boron-nitride,[31] we know that both the TCNQ and TTF are physiorbed on these systems, and they are stabilized at a distance of ~ 3.2 Å over graphene and at a distance of ~ 3.5–3.6 Å over boron-nitride sheet. So, for the optimization calculations, we chose the initial distances of 3.2 Å and 3.5 Å for the dopants on top of the GQDs and BNQDs, respectively. Also, while choosing the initial arrangement of the dopant on top the QDs we followed the references 29 and 31(similar to figures 2 and 3 of this article), respectively, for GQDs and BNQDs. But, to check the universality of these arrangements we have also considered the other configurations, namely, the reported stable arrangement for "TTF on BN-sheet" as an initial arrangement for "TTF on GQD" and the reported stable arrangement for "TCNQ on graphene" as an initial arrangement for "TCNQ on BNQD" etc. as shown in figure **S1**. From these studies we recognized that (see table **S1**) the most stable arrangements of dopants didn't change with dimensionality, for example, they are same for BN-sheet and BNQD.

Next, as both graphene and boron-nitride nanoribbons are shown to be spin-polarized,[38, 39] under different conditions, we have also considered the spin-polarization in all our studies and the energies of the complexes in different spin-states, with respect to their ground-states, are given in table 1. From these spin-polarized calculations, we find that the GQD-dopant complexes are anti-ferromagnetic (AFM) in their ground state, whereas, the BNQD-dopant complexes are nonmagnetic (NM) in their ground state. Next, the stability of each complex in its ground-state has been calculated and the formation energy ($E_{form}$) values of all the (21, 8) complexes are given in the table 2.

Table 2: Formation energy of the complexes, optimum-distance of dopants above QDs and the amount of charge-transfer between dopants and QDs for (21, 8) QDs are given.

| Molecule | $E_{form}$ (eV) | Distance (Å) | Charge-transfer (e) | $E_{form}$ (kcal/mol) |
|---|---|---|---|---|
| GQD + TCNQ | -2.01 | 3.11 | -0.43 | -46.35 |
| GQD + TTF | -1.45 | 3.15 | +0.12 | -33.44 |
| BNQD + TCNQ | -1.66 | 3.38 | -0.18 | -38.28 |
| BNQD + TTF | -1.34 | 3.41 | +0.21 | -30.90 |

Formation energy is calculated using the equation "$E_{form} = E_{complex} - E_{dopant} - E_{QD}$", where $E_{form}$ is the formation energy of the complex and $E_{complex}$, $E_{dopant}$ and $E_{QD}$ are the absolute energies of the complex, dopant and quantum-dot, respectively. It is known that,[29] greater the formation energy of the complex stronger the interaction between the dopant and the substrate (here, QDs). From table 2, it is apparent that the dopants are strongly adsorbed on GQDs than on BNQDs. Presence (absence) of a π-surface through which a GQD (BNQD) can (cannot) interact with its dopant's π-surface could be the main reason for the strong (weak) interaction between GQDs (BNQDs) and the dopants. Values of formation energy are clearly reflected in the distances between the dopants and the QDs – greater the distance lesser is the formation energy – in their optimized structures (see columns 3 and 5 of table 2). In what follows, first we will discuss the properties of the GQD-dopant complexes and then we will switch to the BNQD-dopant complexes.

GQDs + Dopants: Optimized structures of the GQD-dopant complexes, from which the distance between dopants and GQDs are calculated, are given in figure 2. In their minimum energy configurations, TTF's C=C bond is exactly on top of the centre of a benzenoid ring of GQD and TCNQ's benzenoid ring is exactly on top of the benzenoid ring of GQD in staggered position. From the optimized geometries we find that, TCNQ is stable at a distance (this is the least distance between GQD plane and dopant) of ~ 3.11 Å and TTF is stable at a distance of ~ 3.15 Å on top of the (21, 8) GQDs. Also, both the dopants and GQD have bent from their planar structures, but in opposite directions (i.e. dopant has bent in convex manner and GQD in a concave manner). Among the dopants, TCNQ has bent more when compared to TTF (as shown in figures 2a, 2b), and in the GQD, bent has occurred mainly in the area below the dopant.

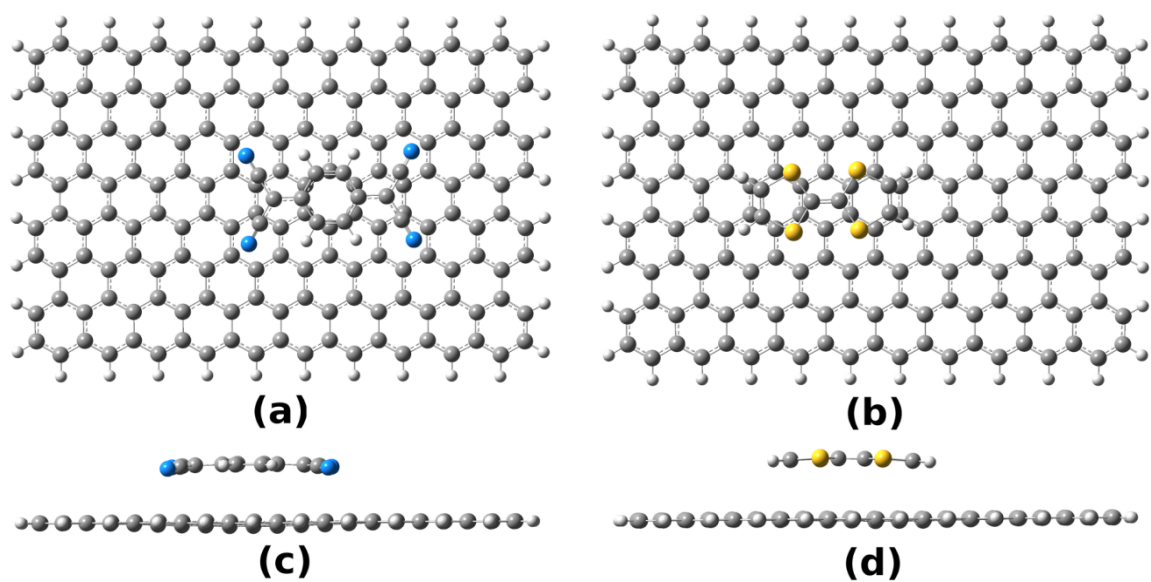

Figure 2: Top row shows the top view of (a) TCNQ and (b) TTF adsorbed on (21, 8)-GQD; Bottom row shows the side view of (c) TCNQ and (d) TTF adsorbed on (21, 8)-GQD

We have also calculated the $E_{form}$ values of the GQD-dopant complexes (see table 2), and we find that these values are comparable (with in 2 kcal/mol) to the $E_{form}$ values of the graphene-dopant complexes,[29] suggesting that the dopants interact with the graphene and GQDs in a similar manner. The lesser $E_{form}$ values (< 50 kcal/mol) also suggest that dopants are just physisorbed on the GQDs. The argument regarding the physisorption is further supported by the large distance of separation – a distance (> 3 Å) at which a chemical bond formation has not been shown till date between a carbon atom and any other atom present in the study[40] – between the dopant and the GQDs (see table 2). Importantly, a larger value of distance separation, a lesser value of bending and a lesser $E_{form}$ value of TTF when compared to TCNQ clearly indicates that there is a strong (weak) interaction between TCNQ (TTF) and GQD.

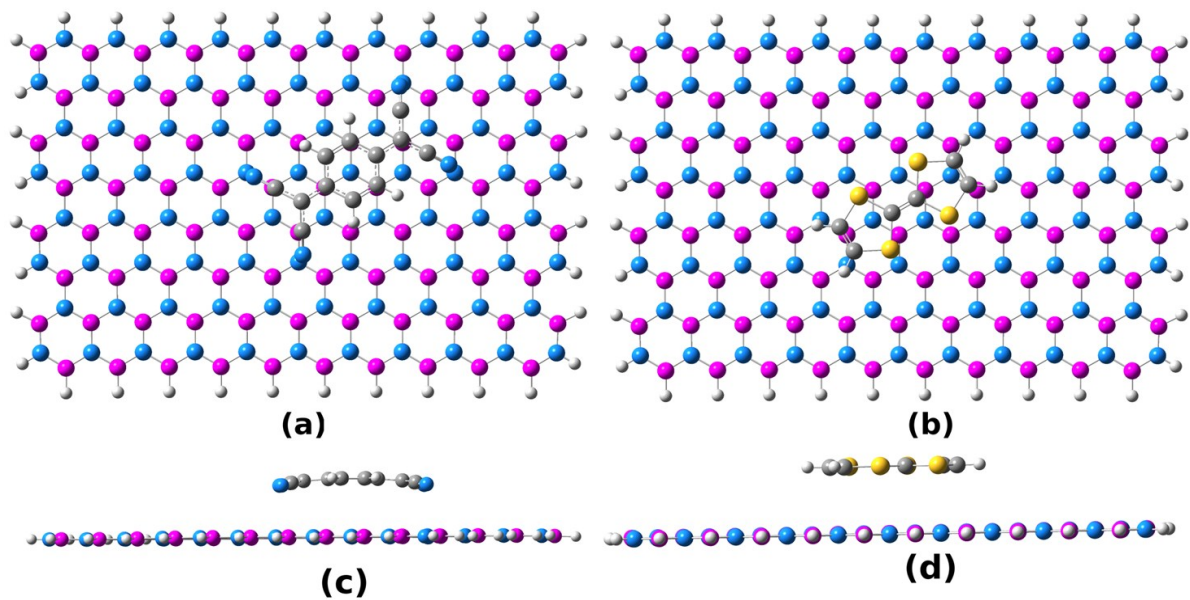

Figure 3: (a) Top-view of TCNQ adsorbed on (21, 8) BNQD and (b) TTF adsorbed on (21, 8) BNQD, (c) side-view of TCNQ adsorbed on (21, 8) BNQD and (d) TTF adsorbed on (21, 8) BNQD

BNQDs + Dopants: Optimized structures of the BNQD-dopant complexes are given in figure 3. The BNQD-TCNQ and BNQD-TTF distances, in their optimized structures, are 3.38 Å and 3.41 Å, respectively. Similar to the GQD-dopant complexes, BNQD-dopant complexes have also deviated from their planarity. Again, TCNQ has bent more compared to the TTF, proving its stronger interaction with BNQD (as shown in figures 3c, 3d). Bending of TCNQ on top of BN-sheet has been observed previously [31] and the reason is predicted as the weak dipolar interaction between the TCNQ and BN-sheet. This is different from the case of graphene-TCNQ, where, the interaction was predicted to be mainly through π-surfaces of dopant and graphene. [29] So, we also expect that the nature of interaction between BNQD and dopants as the weak dipole–dipole and/or electrostatic, similar to its 2-D counterpart.[29, 31] Finally, we noticed that in their optimized structures, TTF's middle C=C bond is on top of the hexagonal ring of the BNQD and TCNQ's benzenoid ring is on top of the BNQD's hexagonal rings in a manner such that three of its six carbon atoms are on top of the hollow sites of the BNQD's hexagonal rings (exactly similar to the Bernald-stacking of graphite[8]). Again, the absence of any chemical bonds between BNQD and the dopants suggests that these molecules are physisorbed.

Table 3: Spin-polarized H-L gaps of TCNQ and TTF adsorbed (21, 8)-QDs. Charge-transfer values are given, again, for comparison.

| System | H-L gap | | Charge-transfer (e) |
|---|---|---|---|
| | α-spin (eV) | β-spin (eV) | |
| Pure GQD | 0.54 | 0.54 | - |
| TCNQ-GQD | *0.21* | *0.09* | -0.43 |
| TTF-GQD | 0.22 | 0.22 | +0.12 |
| Pure BNQD | 4.03 | 4.03 | - |
| TCNQ-BNQD | 0.73 | 0.73 | -0.18 |
| TTF-BNQD | 1.98 | 1.98 | +0.21 |

**Electronic properties**

In this section, first we will explain the charge-transfer between the dopant-QD complexes. Then we will describe the changes in the H-L gap for all the systems, and finally, we will try to explain some of these results based on the density of states (DOS) and projected-DOS (pDOS) plots.

Charge-transfer: In order to find out the amount of charge-transfer between QDs and dopants, we have performed the Mülliken population analysis and this charge-transfer data is given in table 2 for all the (21, 8)-GQD/BNQD-dopant complexes. Population analysis shows that these QDs have the ability to both give and take the charges from the dopants depending on their nature. From table 2 we can also find that, among the QDs, GQD can give electrons to the dopant easily rather than taking from it and exactly the opposite behaviour is exhibited by the BNQD. Now comparing the dopants, there is a significant amount of charge-transfer (see table 2) from GQD to TCNQ and a relatively less amount of charge-transfer from TTF to GQD. The latter is mainly due to the electron-rich nature of the GQD. On the other hand, the electron poor nature of the BNQD could be the reason for the larger charge-transfer from TTF to BNQD compared to the charge-transfer from BNQD to TCNQ.

HOMO–LUMO gap: As mentioned earlier, it is known that the low-dimensional graphene and BN systems can be spin-polarized. So, we have calculated the spin-polarized HOMO–LUMO (H-L) gaps for all the systems and their values for (21, 8)-QD-complexes are given in table 3. From table 3 we can notice that, pure (21, 8)-GQD is a semi-conductor with an H-L gap of 0.54 eV and pure (21, 8)-BNQD is an insulator with an H-L gap of 4.03 eV, for both the spin channels. Interestingly, H-L gap of both the QDs have decreased to more than half of their original H-L gap (though the decrement is huge for BNQDs) with the addition of dopants. In addition to the decrement in the H-L gap, we have also found a spin-dependent H-L gap in the TCNQ-GQD system. But, this spin dependent H-L gap in not observed for the case of TTF. Finally, it is known that the chemical reactivity of a molecule is dependent on the H-L gap in an inverse manner–that is, higher the gap, lower the reactivity.[41] Assuming that the charge-transfer will occur only when a system is chemically reactive, we find that our results on charge-transfer also reflects the same – larger the H-L gap, greater the charge-transfer (please compare columns 2/3 and 4 of table 3) – though, not quantitatively.

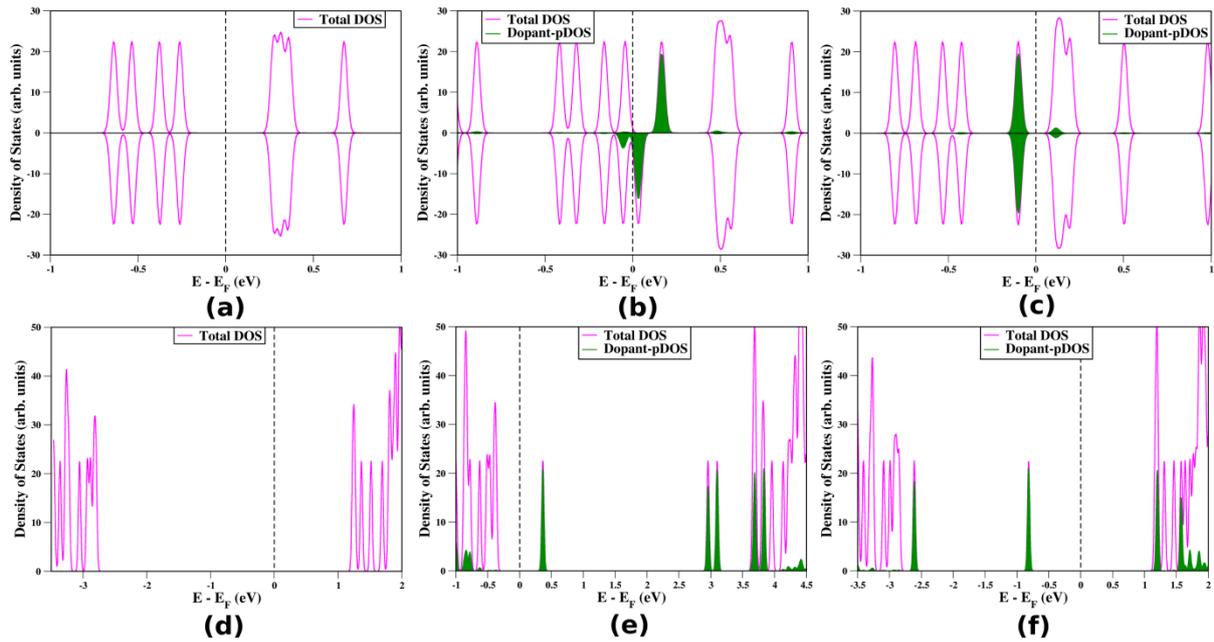

Figure 4: Projected density of states (pDOS) plots of (21, 8) systems. (a) GQD, (b) TCNQ-GQD, (c) TTF-GQD, (d) BNQD, (e) TCNQ-BNQD and (f) TTF-BNQD systems. The Fermi-level is set to zero. Broadening parameter of 0.025 eV is used to broaden the energy levels.

DOS and pDOS: In order to further understand the reasons behind the decrement in the H-L gap in QD-dopant complexes and spin-polarized H-L gap in TCNQ-GQD system we have plotted the DOS and pDOS of all the (21, 8) complexes, as shown in figure 4. From this figure it is apparent that, (a) the DOS of a QD-dopant complex is nearly equal to the sum of the individual DOS of the QD and the dopant and the minor changes are due to the complex formation. This observation shows that the interaction between the dopant and the QD is weak in nature. Also, this proves that the dopant is just physisorbed on QDs; (b) there is a shift in the Fermi-level towards the HOMO or LUMO depending on whether the dopant is an electron-acceptor (TTF) or electron- donor (TCNQ), respectively; (c) major amount of the dopant levels are concentrated near the Fermi-level and are in between the HOMO and LUMO levels corresponding to the un-doped QDs; (d) the position of these major amount of dopant levels is dictated by the electron-accepting/donating nature of the dopant. Thus, for TCNQ – being an electron acceptor – it is just below the LUMO of the QD, and for TTF – being an electron donor – it is just above the HOMO of the QD.

Based on the above observations, we understood that the reason for the decrement in the H-L gap of QDs after the physisorption of a dopant is mainly because of the presence of the dopant-levels near the Fermi-level of the complexes. And, the reason for the generation of spin-dependent H-L gap in the TCNQ-GQD system is due to the spin-polarized TCNQ levels near the Fermi-level. It is important to note that this spin-polarization is due to a combined effect of the GQD and TCNQ and just not because of any one of them. This is because, if GQD (TCNQ) alone can create the spin-polarization irrespective of the dopant (QD), then, obviously, one would expect the same spin-polarization for the TTF-GQD (TCNQ-BNQD) complex. But, none of the above two cases were observed in our study. Furthermore, none of the above molecules – that is, neither the dopants nor the QDs – are spin-polarized individually (see figure S2, give the DOS plots for all the individual molecules). All these above results prove that, there is a combined effect of both GQD and TCNQ for the generation of spin-polarized H-L gap in TCNQ-GQD complex.

Now, as we understood that the spin-polarization of TCNQ levels in the complex is due to the combined effect of both GQD and TCNQ, we have looked at the major changes in the GQD and TCNQ after they formed as a complex. Two such major changes which we observed in the complex are (a) bending of TCNQ/GQD and (b) a change in the total charge on TCNQ/GQD. From table 2, we already know that GQD has transferred ~ 0.43 e to the TCNQ. Further analysis on the charge-transfer shows that, in the transferred ~ 0.43 e charge, ~ 0.045 e has transferred as up-spin and ~ 0.384 e has transferred as down-spin. Thus, we find that the spin-polarized H-L gap of TCNQ-GQD complex has an origin in the spin-polarized charge-transfer from GQD to TCNQ. Also, it is important to note that similar spin-polarized charge-transfer has not been observed for any other complexes of the present study.

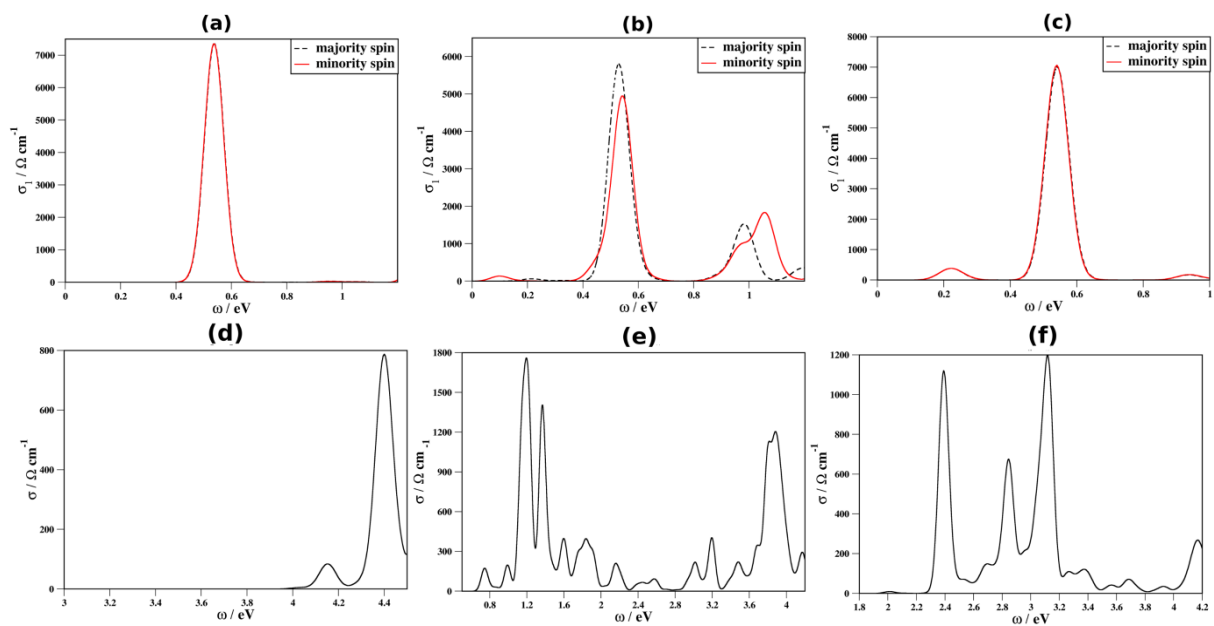

Figure 5: Optical-conductivity profiles (in low-frequency region) of (21, 8) complexes. (a) GQD, (b) TCNQ-GQD, (c) TTF-GQD, (d) BNQD, (e) TCNQ-BNQD, and (f) TTF-BNQD. The dashed black and the solid red lines correspond to the majority and minority-spin optical-conductivities, respectively. The solid black line corresponds to the spin-independent optical conductivity. The lines are broadened with Gaussian functions of width 0.05 eV

**Optical-Conductivity**

It is known that molecular charge-transfer can affect the optical-conductivity in the low-frequency region.[29] As our Mülliken population analysis shows that there is a charge transfer between the dopants and the QDs (see table 3) in the QD-dopant complexes, we have calculated the optical-conductivity of all the (21, 8) complexes and are shown in figure 5. The low-frequency optical-conductivity of GQD is given in figure 5a, and it has a single peak situated at 0.54 eV. This energy is exactly equal to the H-L gap of the GQD (see table 3). Thus, this peak corresponds to the electronic transition from HOMO to LUMO. Also, it should be noted that there no peaks below this energy.

Now, when a dopant is adsorbed on the GQD, new peaks below 0.54 eV are observed. For the TCNQ-GQD complex, these new peaks are situated at 0.21 eV and 0.09 eV (exactly equal to the spin-polarized H-L gaps of TCNQ-GQD complex) for the majority and minority spin-channels, respectively. Adsorption of TCNQ on GQD gives rise to an asymmetry in the population of majority and minority spins near the Fermi-level (see figure 4b) which leads to the difference in conductivity for different spins in the low-frequency regime. But, for the

TTF-GQD complex system there is no spin-polarization, and hence, the optical conductivity is also same for both the spins. For pure BNQD, the calculated H-L gap is 4.03 eV, and for the dopant-BNQD systems, we have found that multiple numbers of peaks appears below 4 eV. These peaks correspond to the electronic excitations between the molecular-molecular and molecular-BNQD energy levels – a clear reflection of the pDOS plots of (21, 8) BNQD-dopant complexes.

Table 4: Spin-polarized H-L gaps of TCNQ and TTF adsorbed on (21, 6), (21, 8) and (33, 8)-QDs.

| Size | Pure-QD H-L gap (eV) | | TCNQ-QD H-L gap (eV) | | TTF-QD H-L gap (eV) | |
|---|---|---|---|---|---|---|
| **GQDs** | Alpha-spin | Beta-spin | Alpha-spin | Beta-spin | Alpha-spin | Beta-spin |
| (21, 6) | 0.68 | 0.68 | 0.22 | 0.11 | 0.24 | 0.23 |
| (21, 8) | 0.54 | 0.54 | 0.21 | 0.09 | 0.21 | 0.22 |
| (33, 8) | 0.54 | 0.54 | 0.21 | 0.06 | 0.18 | 0.18 |
| **BNQDs** | Alpha-spin | Beta-spin | Alpha-spin | Beta-spin | Alpha-spin | Beta-spin |
| (21, 6) | 4.16 | 4.16 | 0.76 | 0.76 | 2.00 | 2.00 |
| (21, 8) | 4.04 | 4.04 | 0.74 | 0.74 | 1.98 | 1.98 |
| (33, 8) | 4.06 | 4.06 | 0.67 | 0.67 | 2.00 | 2.00 |

**Effects of size**

To find the effect of the size of the graphene quantum dots on the interaction between the molecules and the QD, we have considered QDs of two other sizes, namely, (21, 6) and (33, 8). Among them (21, 6) has lesser ribbon width compared to (21, 8), whereas, (33, 8) has larger ribbon length compared to (21, 8). From these calculations we find that, with a change in the size of the QD there are no appreciable changes in either the distance of dopant from QD (changes are less than 0.04 Å) or the amount of charge-transfer between QD and dopant (changes are less than 0.03 e). But, we find that there is a decrement in the H-L gap of the complex with an increase in the size of the QD irrespective of the dopant. The values of H-L gap for different sizes are given in the table 4. The above results clearly show that dopants interact with the QDs in a similar manner irrespective of the QD's size, and these results can also be used to see why there is no significant difference in the dopants interaction with QDs (GQDs and BNQDs) and their 2-D counter parts (Graphene and BN-sheet).

**Conclusions and Outlook:**

In conclusion, we have shown that H-L gap of the QDs can be tuned using organic molecules as dopants, and hence, these materials can be used in opto-electronic devices. Our spin-polarized calculations show that GQD-dopant complexes are stabilized in AFM state and BNQD-dopant complexes in NM state. Among the QDs, GQD has a stronger interaction with dopants compared to BNQD because of its π-surface. Both QDs and dopants are bent (in an opposite manner) in their optimized structures, the latter being more apparent from the respective figures. Among the dopants, TCNQ has interacted strongly with QDs compared to TTF and this has been proved by several factors like lesser QD-dopant distance, larger bending etc. We have noticed that, charge-transfer between the dopant and the QD is inversely proportional to the H-L gap of the complexes, though qualitatively. Cause for the huge decrement in the H-L gaps in QD-dopant complexes compared to the pure QDs has been explained using the pDOS plots–main reason being the presence of the molecular-levels near the Fermi-energy. And, the origin for the spin-polarized H-L gap of the TCNQ-GQD complex had been found in the spin-polarized charge transfer from GQD to TCNQ. Optical conductivities of QDs have changed with doping and these changes have been understood based on the changes in H-L gap and pDOS plots. Finally, the size variation studies of QDs show that size has a major impact only on the H-L gap.

**Notes and References:**

*a. New Chemistry Unit, Jawaharlal Nehru Centre for Advanced Scientific Research, Jakkur P.O., Bangalore – 560064, India.*

*b. Chemistry and Physics of Materials Unit, Jawaharlal Nehru Centre for Advanced Scientific Research, Jakkur P.O., Bangalore – 560064, India.*

*c. Theoretical Sciences Unit, Jawaharlal Nehru Centre for Advanced Scientific Research, Jakkur P.O., Bangalore – 560064, India.*

*‡ These authors have contributed equally to this work*

\* Corresponding author; E-mail: pati@jncasr.ac.in

**† Electronic Supplementary Information (ESI) available:** Details of (a) structures and energies of QD-dopant complexes in different conformations; (b) DOS plots of GQD, TTF and TCNQ.

# Supporting Information

Table **S1**: Energies of the GQD-dopant complexes with a variation in the position of the dopant on QD. Energies are scaled to the most stable conformation.

| QD-dopant Complex | Energy of the structure in Figure 2 (eV) | Energy of the structure in Figure S1 (eV) |
|---|---|---|
| GQD-TCNQ | 0.00 | 0.06 |
| GQD-TTF | 0.00 | 0.04 |
| BNQD-TCNQ | 0.00 | 0.05 |
| BNQD-TTF | 0.00 | 0.04 |

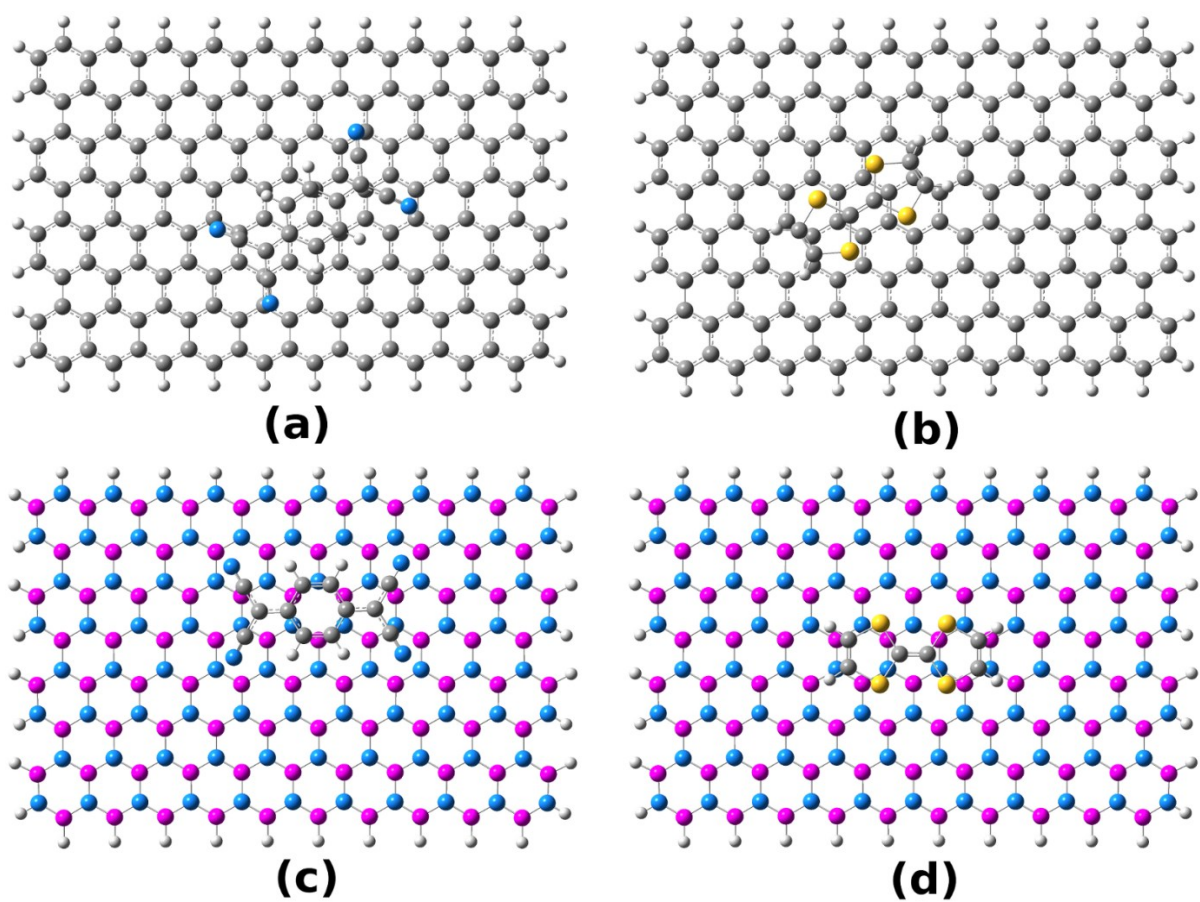

Figure **S1**: (a) TCNQ adsorbed on (21, 8) GQD, (b) TTF adsorbed on (21, 8) GQD, (c) TCNQ adsorbed on (21, 8) GQD and (d) TTF adsorbed on (21, 8) GQD. In this figure, the positions of the TCNQ and TTF on GQD are swapped with their positions on BNQD in the figure 2, and vice-a-versa for BNQD.

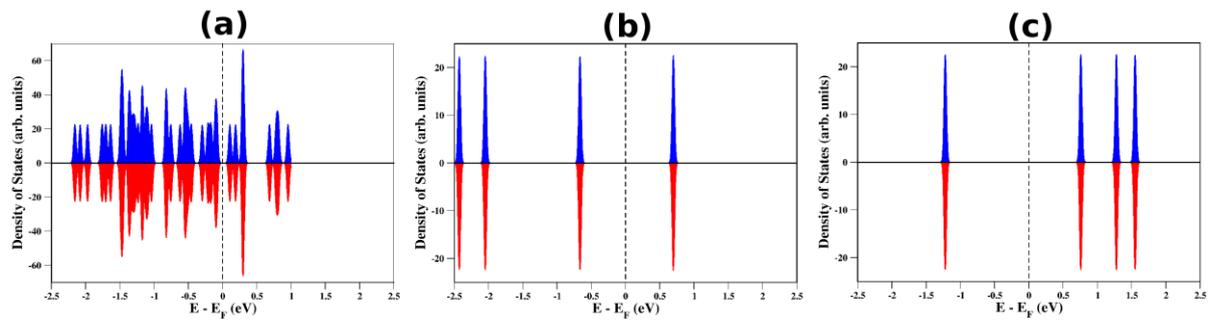

Figure **S2**: DOS plots of (a) (21, 8)-GQD, (b) TCNQ and (c) TTF show all the systems are non-spin-polarized.